\documentclass[aps,prd,reprint,superscriptaddress,nofootinbib,floatfix,longbibliography]{revtex4-1}
\usepackage[dvipsnames]{xcolor}
\usepackage{amssymb,amsmath,amsfonts,bm,slashed,soul,ragged2e,graphicx,epstopdf,hyperref,array}
\usepackage[utf8]{inputenc}
\usepackage{caption,subcaption}
\enlargethispage{\baselineskip}
\hypersetup{colorlinks,linkcolor={blue},citecolor={blue},urlcolor={blue}}
\DeclareCaptionJustification{justified}{\justifying}
\everymath{\displaystyle} 

\begin{document}

\title{Superradiant evolution of the shadow and photon ring of Sgr A$^\star$}

\author{Yifan Chen}
\email{yifan.chen@nbi.ku.dk}
\affiliation{CAS Key Laboratory of Theoretical Physics, Institute of Theoretical Physics,\\
Chinese Academy of Sciences, 100190 Beijing, People's Republic of China}

\author{Rittick Roy}
\email{rittickrr@gmail.com}
\affiliation{Center for Field Theory and Particle Physics and Department of Physics, Fudan University, 200438 Shanghai, People's Republic of China}

\author{Sunny Vagnozzi}
\email{sunny.vagnozzi@ast.cam.ac.uk}
\affiliation{Kavli Institute for Cosmology (KICC) and Institute of Astronomy,\\University of Cambridge, Madingley Road, Cambridge CB3 0HA, United Kingdom}

\author{Luca Visinelli}
\email{luca.visinelli@sjtu.edu.cn}
\affiliation{Tsung-Dao Lee Institute (TDLI), 520 Shengrong Road, 201210 Shanghai, People's Republic of China}
\affiliation{School of Physics and Astronomy, Shanghai Jiao Tong University, 800 Dongchuan Road, 200240 Shanghai, People's Republic of China}

\date{\today}

\begin{abstract}
\noindent Ultralight bosons can affect the dynamics of spinning black holes (BHs) via superradiant instability, which can lead to a time evolution of the supermassive BH shadow. We study prospects for witnessing the superradiance-induced BH shadow evolution, considering ultralight vector and tensor fields. We introduce two observables sensitive to the shadow time-evolution: the shadow drift, and the variation in the azimuthal angle lapse associated to the photon ring autocorrelation. The two observables are shown to be highly complementary, depending on the observer's inclination angle. Focusing on the supermassive object Sgr A$^\star$ we show that both observables can vary appreciably over human timescales of a few years in the presence of superradiant instability, leading to signatures which are well within the reach of the Event Horizon Telescope for realistic observation times (but benefiting significantly from extended observation periods), and paving the way towards probing ultralight bosons in the $\sim 10^{-17}\,{\rm eV}$ mass range.
\end{abstract}

\maketitle

\section{Introduction}

Signatures of the presence and dynamics of black holes (BHs) are now regularly observed by a diverse range of probes, turning these peculiar regions of spacetime into extreme laboratories for testing fundamental physics~\cite{Barack:2018yly}. Very long baseline interferometry (VLBI) has unlocked the possibility of imaging the shadows of supermassive BHs (SMBHs), i.e.\ their dark silhouette against the imprint of the surrounding radiation~\cite{Luminet:1979nyg,Falcke:1999pj,Perlick:2021aok}. In 2019 the Event Horizon Telescope (EHT) delivered the first groundbreaking horizon-scale images of the shadow of M87$^\star$, the SMBH situated at the center of the nearby elliptical galaxy Messier 87$^\star$~\cite{EventHorizonTelescope:2019dse, EventHorizonTelescope:2019uob, EventHorizonTelescope:2019jan, EventHorizonTelescope:2019ths, EventHorizonTelescope:2019pgp, EventHorizonTelescope:2019ggy, EventHorizonTelescope:2021bee, EventHorizonTelescope:2021srq}.

These results were recently followed by the EHT's first image of the shadow of Sagittarius A$^\star$ (Sgr A$^\star$), a bright and compact radio source residing in our Galactic Center~\cite{Ghez:2008ms}, possibly identified as a SMBH of mass $M_{\rm Sgr\,A^\star} = (4.154\pm 0.014)\times 10^6\,M_\odot$ (calibrated by astrometric tracing of the orbiting S-stars) at a distance $d = (8178\pm 13\pm 22)\,{\rm pc}$ from us~\cite{GRAVITY:2018ofz, 2019A&A...625L..10G,Do:2019txf}. Sgr A$^\star$'s short dynamical timescale $\sim 20\,$s prompted the EHT to image its dynamical evolution by reconstructing the source's emission region~\cite{2017ApJ...850..172J, 8361036}. Assuming general relativity, the estimated dynamical mass and distance are consistent with the size of the bright ring of emission $(51.8 \pm 2.3)\,\mu{\rm as}$ reported by the EHT~\cite{EventHorizonTelescope:2022xnr, EventHorizonTelescope:2022vjs, EventHorizonTelescope:2022wok, EventHorizonTelescope:2022exc, EventHorizonTelescope:2022urf, EventHorizonTelescope:2022xqj, EventHorizonTelescope:2022gsd, EventHorizonTelescope:2022ago, EventHorizonTelescope:2022okn, EventHorizonTelescope:2022tzy}.

VLBI horizon-scale images are paving the way for tests of fundamental physics in the strong gravity regime (see e.g.\ Refs.~\cite{Held:2019xde,Wei:2019pjf,Shaikh:2019fpu,Jusufi:2019nrn,Vagnozzi:2019apd,Long:2019nox,Zhu:2019ura,Contreras:2019cmf,Qi:2019zdk,Neves:2019lio,Javed:2019rrg,Banerjee:2019nnj,Shaikh:2019hbm,Kumar:2019pjp,Allahyari:2019jqz,Li:2019lsm,Jusufi:2019ltj,Rummel:2019ads,Kumar:2020hgm,Vagnozzi:2020quf,Li:2020drn,Narang:2020bgo,Liu:2020ola,Konoplya:2020bxa,Guo:2020zmf,Pantig:2020uhp,Wei:2020ght,Kumar:2020owy,Islam:2020xmy,Chen:2020aix,Sau:2020xau,Jusufi:2020dhz,Kumar:2020oqp,Chen:2020qyp,Zeng:2020dco,Neves:2020doc,Ovgun:2020gjz,Badia:2020pnh,Jusufi:2020cpn,Khodadi:2020jij,Belhaj:2020nqy,Jusufi:2020agr,Kumar:2020yem,Jusufi:2020odz,Belhaj:2020mlv,Kruglov:2020tes,EventHorizonTelescope:2020qrl,Ghasemi-Nodehi:2020oiz,Ghosh:2020spb,Khodadi:2020gns,Lee:2021sws,Contreras:2021yxe,Shaikh:2021yux,Afrin:2021imp,Addazi:2021pty,EventHorizonTelescope:2021dqv,Shaikh:2021cvl,Badia:2021kpk,Wang:2021irh,Khodadi:2021gbc,Cai:2021uov,Frion:2021jse,Wei:2021lku,Rahaman:2021web,Walia:2021emv,Afrin:2021wlj,Jusufi:2021fek,Cimdiker:2021cpz,Pal:2021nul,Li:2021ypw,Afrin:2021ggx,Pantig:2022toh,He:2022yse,Jha:2022ewi,Meng:2022kjs} for M87$^\star$ and Refs.~\cite{Chen:2022lct,Jusufi:2022loj,Vagnozzi:2022moj,Uniyal:2022vdu,Vagnozzi:2022tba,Carballo-Rubio:2022imz,Pantig:2022ely,Ghosh:2022kit,Kuang:2022ojj,Wang:2022fgj,Khodadi:2022pqh,Banerjee:2022iok,Shaikh:2022ivr} for Sgr A$^\star$), including the possibility that astrophysical BHs may be ``mimickers'', i.e.\ (possibly horizonless) compact objects other than BHs~\cite{Bambi:2008jg,Ohgami:2015nra,Shaikh:2018kfv,Cardoso:2019rvt,Bambi:2019tjh,Joshi:2020tlq,Guo:2020tgv,Herdeiro:2021lwl,Visinelli:2021uve,Solanki:2021mkt,Saurabh:2022jjv,Cardoso:2022fbq,Tahelyani:2022uxw,Patel:2022jbk,Patel:2022vlu}. Here, we shall pursue a similar direction, by characterizing detectable imprints of new ultralight bosons on SMBH shadows, driven by the radiation-enhancement phenomenon of BH superradiance (BHSR)~\cite{Dicke:1954zz,Penrose:1971uk,1971JETPL..14..180Z,Damour:1976kh,Ternov:1978gq,Zouros:1979iw,Detweiler:1980uk}.

New ultralight bosonic fields are well-motivated by a wide range of scenarios, extending from string compactifications to models of dark matter and (early) dark energy~\cite{Peccei:1977hh,Weinberg:1977ma,Wilczek:1977pj,Witten:1984dg,Wetterich:1987fm,Ratra:1987rm,Hu:2000ke,Khoury:2003rn,Svrcek:2006yi,Cicoli:2012sz,Foot:2014uba,Burrage:2016bwy,Hui:2016ltb,Visinelli:2018utg,Poulin:2018cxd,Niedermann:2019olb,Sakstein:2019fmf,DiLuzio:2020wdo,Vagnozzi:2021gjh,Choi:2021aze}. BHSR drives the growth of a cloud of massive bosonic fields orbiting a spinning BH at the expense its energy and angular momentum~\cite{Arvanitaki:2009fg}, with a key role played by the existence of an event horizon, acting as an absorbing boundary condition (see Refs.~\cite{Cardoso:2004nk,Cardoso:2012zn,Herdeiro:2013pia,East:2013mfa,Brito:2014nja,Rosa:2015hoa,Cardoso:2015zqa,Wang:2015fgp,Rosa:2016bli,Degollado:2018ypf,Ikeda:2018nhb,Boskovic:2018lkj,Baumann:2019eav,Cardoso:2020hca,Blas:2020kaa,Baryakhtar:2020gao,Unal:2020jiy,Franzin:2021kvj,Cannizzaro:2021zbp,Jiang:2021whw,Khodadi:2021mct} for important earlier work). BHSR is most efficient when the boson's Compton wavelength is comparable to the BH gravitational radius $r_g$ (crucially relying only on gravitational interactions), and its signatures have been used to constrain new light bosons from observations of spinning BHs~\cite{Arvanitaki:2010sy,Pani:2012vp,Brito:2013wya,Cardoso:2016olt,Cardoso:2018tly,Stott:2018opm,Stott:2020gjj}. As the appearance of a BH shadow is very sensitive to the BH's mass and spin, the BHSR-driven growth of the bosonic cloud growth at the expense thereof leads to an \textit{evolution} of the BH shadow~\cite{Roy:2019esk,Creci:2020mfg,Roy:2021uye}, which in the case of scalar (spin-0) BHSR was argued to be potentially observable~\cite{Roy:2019esk, Roy:2021uye}.

In this paper, we study BHSR-driven signatures which are well within the reach of the EHT and future VLBI arrays, and can lead to a clear detection of the imprint of new light bosons on the shadow of Sgr A$^\star$ within a timescale of a few years or decades. More precisely, we consider the imprint of the superradiant evolution as potentially observable through the \textit{i}) shadow drift (defined later), and the \textit{ii}) photon ring autocorrelation. We consider vector and tensor fields, which lead to stronger signatures compared to scalars. Alongside related efforts~\cite{Davoudiasl:2019nlo,Chen:2019fsq,Bar:2019pnz,Cunha:2019ikd,Chen:2021lvo}, our results provide a novel pathway towards probing new ultralight bosons through BH shadows. We set $\hbar=c=1$.

\section{Superradiant evolution}

We consider the evolution over the background of a rotating (Kerr) BH of a vector field $A^\mu$ with mass $\mu_V$ as described by the Proca equation, or alternatively a tensor field $H_{\mu\nu}$ with mass $\mu_T$ as described by Fierz-Pauli equation~\cite{Vainshtein:1972sx,Boulware:1972yco,Hassan:2011hr,Hassan:2011zd,vonStrauss:2011mq,Babichev:2013usa,Babichev:2016bxi}. We refer to the boson mass as $\mu$ when discussing general properties of the boson cloud. On the Kerr background and in spherical coordinates, the wave function for a boson of spin $s$ separates into radial and angular components. The angular part is described by spin-weighted spherical harmonics, with the orbital angular momentum $\ell$ and the azimuthal number $m$ describing the projection of the total angular momentum $j$ along the $z-$axis.

A massive boson field incident upon a spinning BH can develop superradiant instability, leading to exponential increase in the boson occupation numbers at the expense of the BH mass and spin~\cite{Arvanitaki:2009fg,Brito:2015oca}, provided the angular phase velocity of the incident wave $\omega$ satisfies $m\Omega_H > \omega$, where $\Omega_H$ is the angular velocity at the BH event horizon. Given a BH of mass $M$ and in the limit of small gravitational coupling $\alpha \equiv GM\mu$, superradiance is most effective for $\alpha \sim {\cal O}(0.1)$. The superradiant instability rates are given by~\cite{Brito:2015oca,East:2017ovw,Brito:2020lup}
\begin{eqnarray}
	\label{eq:timescale}
	\Gamma_{j\ell} &\simeq& C_{\ell S}\frac{\mathcal{P}_{\ell m}(a)}{\mathcal{P}_{\ell m}(0)}\alpha^{4\ell+2S+5}\left(m\Omega_{\rm BH} \!-\! \omega\right)\,,\\
	\mathcal{P}_{\ell m}(a) &\equiv& (1\!+\!\Delta)\!\prod_{q=1}^\ell\!\!\left[\!\Delta^2\!\!+\!4r_g^2(1+\Delta)^2\!\left(m\Omega_{\rm BH} \!-\! \omega\right)^2\!\right]\!\!,
\end{eqnarray}
where $\Delta \equiv \sqrt{1-a^2}$, with $a$ being the dimensionless spin, $S=\{-s,-s+1,...s-1, s\}$ is the spin projection along the $z$-axis, and the coefficients $C_{\ell S}$ relevant for us are $C_{1-1}=16$ and $C_{2-2} = 128/45$. For Sgr A$^\star$, the associated instability timescale depends on the nature and quantum numbers of the boson cloud, with vector fields extracting energy on a significantly shorter timescale than scalar fields. For example, a vector field with $\alpha = 0.2$ reduces the superradiant timescale for an extremal SMBH to $t \sim \mathcal{O}(10){\rm \,yrs}$, compared to $t \sim \mathcal{O}(10^4){\rm \,yrs}$ for a scalar field with the same $\alpha$. For tensor fields, the existence of two unstable modes with $m=1$ (dipole) and $m=2$ (quadrupole) leads to the quadrupole mode still being in the superradiant phase for $\omega/2 < \Omega_H < \omega$, even after the dipole mode has stopped growing~\cite{Brito:2020lup}, similarly to the case of two scalar fields~\cite{Roy:2021uye} or multiple modes~\cite{Ficarra:2018rfu}.

\subsection{Large inclination angle: Shadow drift}

A change in the BH mass and spin due to BHSR affects the shape, position, and size of the BH shadow. Detecting a change in the features of the BH shadow over a relatively short period of a few years could therefore hint at the existence of new ultralight bosons. 

As the BH's position on the celestial sphere is fixed, we choose it as the origin of the sky plane coordinates. For a given inclination angle and spin, we determine the shadow contour, along which the points with the largest and smallest radial distances from the BH's center lie on the axis perpendicular to the BH spin projection. We refer to the distance between these points and the BH's center as $r_{\max}$ and $r_{\min}$, respectively. Because of this asymmetry, the shadow contour's center does not correspond to the BH position, and the radial distance between the shadow's and BH center is parametrized as
\begin{equation}
r_C \equiv |r_{\max} - r_{\min}|/2\,.\label{eq:rc}
\end{equation}
A change in the BH spin leads to a shift in $r_C$ along the axis perpendicular to the BH spin projection. Alternative definitions, including the radial distance to the shadow's geometrical mean or the ring center defined by the EHT~\cite{EventHorizonTelescope:2019ths,EventHorizonTelescope:2022wok,EventHorizonTelescope:2022exc}, lead to quantitatively similar results. On the other hand, the BH spin evolution has a less significant impact on the overall shape of the shadow contour calculated with respect to the shadow center instead of the BH center, as well as other quantities such as the average shadow radius, the ring diameter, and the circularity. Therefore, in the following, we focus on the drift of the shadow center $r_C$ defined in Eq.~\eqref{eq:rc}. Such a definition benefits from encompassing both shape distortion and position shift in the BH shadow.

\begin{figure}[htb!]
    \includegraphics[width=0.43\textwidth]{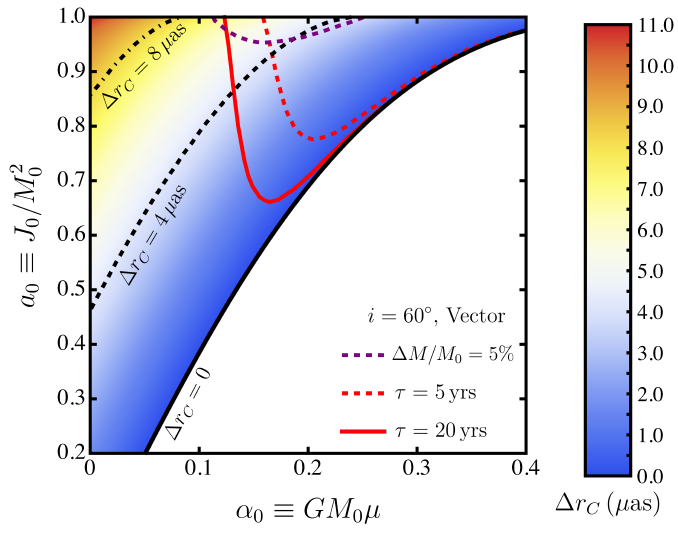}
    \includegraphics[width=0.43\textwidth]{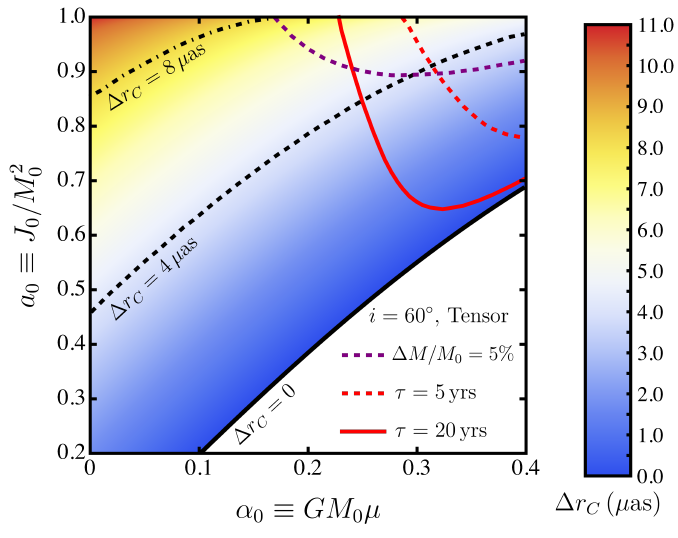}
    \caption{Change in the drift parameter $r_C \equiv |r_{\max} - r_{\min}|/2$ of Sgr A$^\star$'s shadow for different choices of $\alpha_0 \equiv GM_0\mu$ (horizontal axis) and the initial spin parameter $a_0 \equiv J_0/M_0^2$ (vertical axis), for the case of a vector field (top panel) or the quadrupole mode of a tensor field (bottom panel), for inclination angle $i=60^\circ$. The black curves mark an angular change $\Delta r_C = 4{\rm\,\mu as}$ (dashed curve) or $\Delta r_C = 8{\rm\,\mu as}$ (dotted curve), and the minimum value of $a$ for a given $\alpha_0$, corresponding to $\Delta r_C = 0$ (solid curve). Also shown are contours above which the superradiant timescale is below $20\,$yrs (red solid curve) or $5\,$yrs (red dashed curve), and above which the relative variation in the BH mass is $>5\%$ (purple dashed curve).}
    \label{fig:SgrAdensity}
\end{figure}

We consider the superradiant evolution of a boson cloud around Sgr A$^\star$ with initial spin parameter $a_0 \equiv |{\bf J}_0|/M_0^2$, where ${\bf J}_0$ and $M_0$ are the initial spin and mass of the SMBH. We solve for the evolution of the BH spin and mass, as well as for the occupancy number of bosons in the cloud, assuming these quantities evolve over the timescale in Eq.~\eqref{eq:timescale} and setting $M_0 = M_{\rm Sgr\,A^\star}$. We neglect the effects of GW emission and gas accretion, which operate on much longer timescales~\cite{Yoshino:2013ofa,Brito:2014wla,Barausse:2014tra}, as well as the additional assumption that the nonlinear backreaction effects of the bosonic field on the metric are not relevant. We have not included possible effects that would lead to a saturation of the cloud growth, including plasma effects~\cite{Dima:2020rzg}, interactions of the boson field with SM gauge bosons or particles in a dark sector~\cite{Caputo:2021efm}, or self-interactions~\cite{Fukuda:2019ewf,Baryakhtar:2020gao,East:2022ppo}. The initial mass of the boson cloud has little impact on the timescale for most of the parameter space, and we set it to $M_B = M_\odot$, although lower values yield similar results. For a given choice of the initial value of $\alpha_0 \equiv GM_0\mu$, the BH spins down to a final value $a_F$ at which BHSR shuts down, obtained by setting $\omega = m\Omega_{\rm BH}$.

Consensus is lacking regarding the spin and inclination angle of Sgr A$^\star$. The EHT observations of Sgr A$^\star$ are consistent with a large spin and low inclination angle~\cite{EventHorizonTelescope:2022xnr,EventHorizonTelescope:2022urf}, but have not ruled out the antipodal region of parameter space. Earlier work based on quasi-periodic oscillations of emissions in radio, infrared, and x rays~\cite{2010MNRAS.403L..74K} or millimeter VLBI~\cite{Broderick:2008sp, Broderick:2011mk}, exclude an extremal spin. Studies using semianalytical models have reported measurements ranging from $a\lesssim 0.9$ to $a \lesssim 0.5$, while magnetohydrodynamics simulations obtained $a\sim 0.5$~\cite{Shcherbakov:2010ki} or $a\sim 0.9$~\cite{Huang:2009sq}. Finally, Ref.~\cite{Fragione:2020khu} used S-stars motion to set the limit $\vert a \vert \lesssim 0.1$, significantly tighter than the limit obtained considering the surrounding flare emissions~\cite{Matsumoto:2020wul}. In the future, measurements of the twisted light of Sgr A$^\star$ could allow for a direct measurement of its spin~\cite{Tamburini:2019vrf,Tamburini:2021lyi}. Since there is overall no agreement regarding Sgr A$^\star$'s spin and inclination angle, we do not fix the initial spin and perform a scan over all possible spins, within the Kerr limit $a \leq 1$, when presenting the results below in Fig.~\ref{fig:SgrAdensity}. The only requirement is that for a given $\alpha_0$ and $m$, thus a given $a_F$, $a_0 > a_F$ is satisfied.

For a given choice of the parameters $\{a_0, \alpha_0\}$ and for an inclination angle $i$, the BH drift parameter $\Delta r_C$ is defined as:
\begin{equation}
    \label{eq:shadowdrift}
    \Delta r_C \equiv r_C[a_F(\alpha_0), M_F(\alpha_0), i] - r_C(a_0, M_0, i)\,,    
\end{equation}
where the final BH mass $M_F$ is fixed by the properties of the boson and the initial conditions for the BH spin and mass. Values of $\Delta r_C$ in units of $\mu{\rm as}$ are shown in Fig.~\ref{fig:SgrAdensity} for a vector field (top panel) and a tensor field (bottom panel). For illustrative purposes, the inclination angle is $i=60^\circ$. Along the black solid curve $\omega = m\Omega_{\rm BH}$ is satisfied, while dotted and dot-dashed curves mark contours for $\Delta r_C = 4\,\mu{\rm as}$ and $\Delta r_C = 8\,\mu{\rm as}$ respectively. For parameters above the red curves, the superradiance timescale is shorter than 5 years (dashed) or 20 years (solid). The purple dashed line marks the region above which the relative change in the BH mass is larger than 5\%, therefore backreaction on the Kerr metric can start to be relevant, making a numerical approach accounting for this desirable~\cite{East:2017mrj}. The results for the vector case show a region for $\alpha_0 \sim 0.15$ and $a \gtrsim 0.9$ where the shadow drift could be within reach of the EHT along human timescales. The resolution to the shadow center increase significantly for longer observation times, making detection prospects with EHT and the next-generation EHT (ngEHT)~\cite{2021ApJS..253....5R,Lngeht} very optimistic.

\begin{figure}[htb]
    \includegraphics[width=0.45\textwidth]{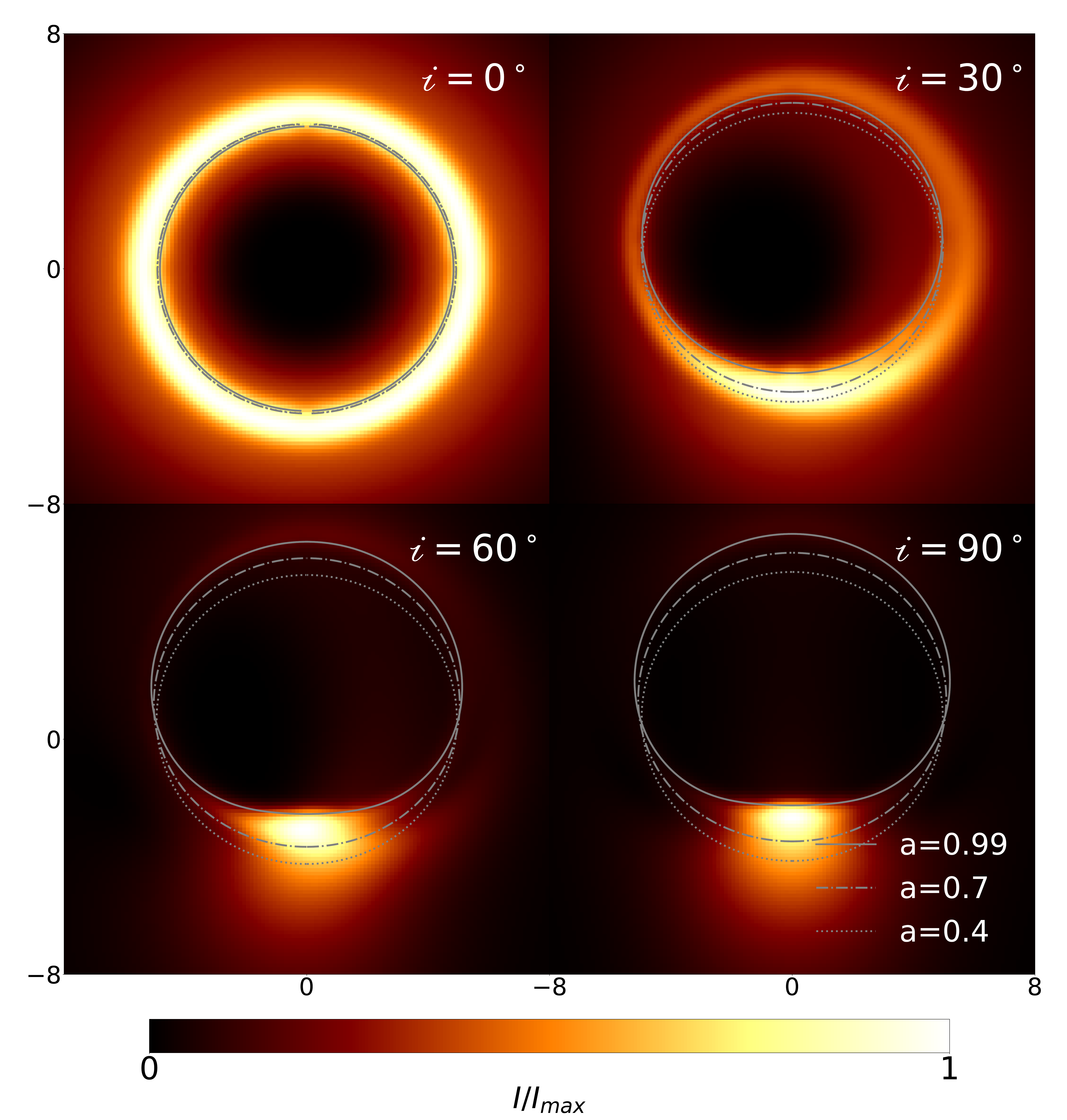}
    \caption{Background images of normalized intensity maps for a fast-rotating SMBH with $a_0=0.99$ at different inclination angles $i$. The gray lines represent the shadow contour at different stages of the evolution for a vector boson with $\alpha_0 = 0.1$. Coordinate axes are in units of $r_g$ at the initial time, with the origin being the BH center and the spin projecting in the horizontal direction.} 
    \label{fig:SgrA_images}
\end{figure}

Figure~\ref{fig:SgrA_images} shows several examples of the shadow evolution for a vector boson with $\alpha_0 = 0.1$, at different inclination angles $i$. The background normalized intensity ($I/I_{\max}$) maps are generated using the radiative transfer code \texttt{IPOLE}~\cite{Moscibrodzka:2017lcu, Noble:2007zx} with an analytic radiative inefficient accretion flow model~\cite{Pu:2018ute}. Gray lines represent the evolution of shadow contours at different stages of the superradiant evolution. Since the shadow drift increases with $i$, this signature is particularly promising for high inclination angles.

\subsection{Low inclination angle: Photon ring autocorrelation}

So far, we have discussed the change in the angular size of the shadow drift for the cases of a relatively large inclination angle, fixing $i = 60^\circ$ for illustrative purposes. However, Fig.~\ref{fig:SgrA_images} shows that the shadow drift decreases at lower inclination angles and vanishes for the face-on case. In this regime, the spin evolution is best probed using the photon ring autocorrelation~\cite{Hadar:2020fda}.

Due to the strong gravitational field outside the BH, photons can orbit around the BH multiple times before reaching the celestial plane, locally enhancing the intensity with a narrow ringlike structure~\cite{Johannsen:2010ru, Gralla:2019xty, Johnson:2019ljv, Gralla:2019drh}. For each emission, light rays with different numbers of half-orbits around the BH finally arrive at different points on the celestial plane at different times. Intensity fluctuations $\Delta I$ along the ring are thus correlated for points separated by a certain azimuthal angle lapse $\delta_0$ and a time delay $\tau_0$. More explicitly, starting from the two-point correlation function of intensity fluctuations and integrating out the direction perpendicular to the azimuthal direction of the ring leads to the function:
\begin{equation}
    \label{eq:intensity}
    \mathcal{C}(T, \Phi)\equiv\iint \mathrm{d} r \mathrm{d} r^{\prime} r\,r^\prime \left\langle\Delta I(t, r, \phi) \Delta I\left(t\!+\!T, r^{\prime}, \phi\!+\!\Phi\right)\right\rangle,
\end{equation}
where $(r, \phi)$ represents the polar coordinate of the celestial plane and $t$ is the observation time. The function $\mathcal{C}(T, \Phi)$ in Eq.~\eqref{eq:intensity} only depends on the time delay $T$ and the azimuthal lapse $\Phi$. The function peaks at $T = \tau_0$ and $\Phi = \delta_0$~\cite{Hadar:2020fda}, where the critical parameters $\tau_0$ and $\delta_0$ describe the additional time and azimuthal lapse spent by the photon traveling one more half-orbit from the same source and depend only on the external space-time properties. A measurement of the critical parameters using the photon ring autocorrelations in Eq.~\eqref{eq:intensity} can provide precise information about the BH. Due to the frame-dragging effect, the azimuthal angle lapse $\delta_0$ is strongly sensitive to the BH spin, ranging from $271^\circ$ for extremal BHs to $180^\circ$ for nonrotating ones~\cite{2003GReGr..35.1909T, Gralla:2019drh}.

Within the current EHT setup, about one year of observation time can be sufficient to reach the sensitivity required to resolve the photon ring surrounding Sgr A$^\star$'s shadow~\cite{Hadar:2020fda}, allowing to measure $\delta_0$ and access information on the spin decrease throughout a $\mathcal{O}(10){\rm \,yrs}$ superradiant phase. These signatures can of course be observed at much higher significance with ngEHT.

\begin{figure}[htb!]
    \includegraphics[width=0.43\textwidth]{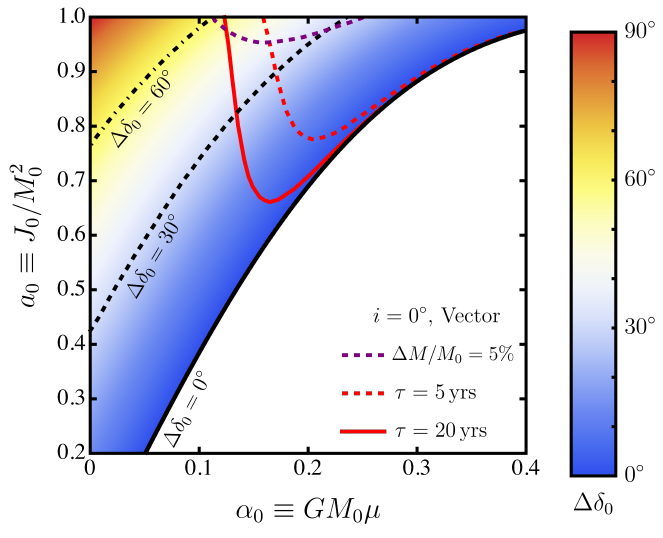}
    \includegraphics[width=0.43\textwidth]{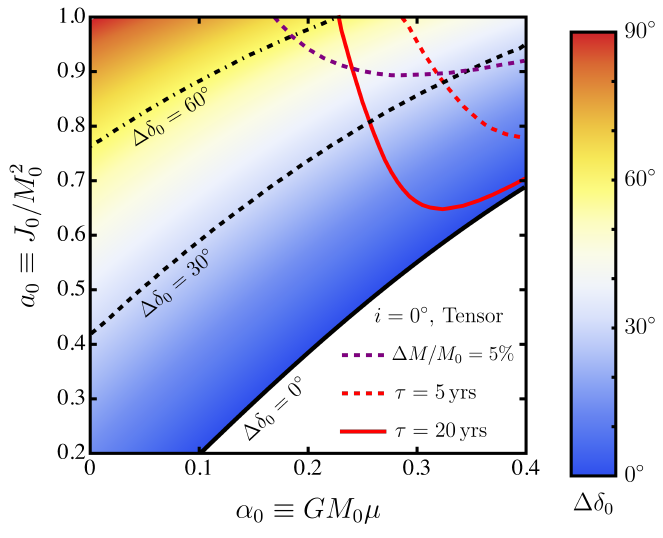}
    \caption{Same as in Fig.~\ref{fig:SgrAdensity} for the variation in the azimuthal angle lapse $\delta_0$ with an inclination angle $i=0^\circ$.}
    \label{fig:SgrAdensity_delta0}
\end{figure}

Figure~\ref{fig:SgrAdensity_delta0} shows variations in the azimuthal angle lapse, $\Delta \delta_0$, as a function of the initial spin $a_0$ and $\alpha_0$ for both vector and tensor fields and for the face-on case $i=0^\circ$. For most of the parameter space, $\Delta \delta_0$ is much larger than the azimuthal correlation length $\ell_\phi \approx 4.3^\circ$ associated to the accretion flow~\cite{Hadar:2020fda}, which sets a theoretical uncertainty floor for $\delta_0$. Larger inclination angles lead to $\phi$-dependent critical parameters $\delta$ whose evolution in terms of spin varies with $\phi$, but remains quantitatively of the same order as in the face-on case~\cite{Gralla:2019drh}.

\section{Discussion and conclusions}

We have explored the possibility of observing the time-evolution of a SMBH shadow induced by superradiance in the presence of ultralight vector and tensor fields. We have shown that the evolution will be accessible through observations of the shadow drift parameter $\Delta r_C$ introduced in Eq.~\eqref{eq:shadowdrift}, and the azimuthal angle lapse $\delta_0$ related to the photon ring autocorrelation in Eq.~\eqref{eq:intensity}.

Variations in the above observables are already accessible by the EHT (and at higher significance by ngEHT), and can take place on human timescales, depending on the nature of the boson species (with vector and tensor superradiance proceeding at a faster rate relative to the scalar case). These two observables are highly complementary. At high inclination angles, the shadow drift is the most promising observables. At low inclination angles where the shadow drift decreases significantly, the azimuthal angle lapse is very sensitive to the SMBH spin. The sensitivity of both observables would significantly benefit from an extended observation time. The angular resolution of VLBI arrays will be further improved by the extended array and higher observation frequencies available with ngEHT~\cite{Blackburn:2019bly, 2021ApJS..253....5R,Lngeht}, allowing to cover a much wider range of parameter space.

For the SMBH Sgr A$^\star$, superradiance would be effective if a new boson of mass $\mu \sim 10^{-17}\,$eV exists. If a time-variation in either of the two observables described above is reported, the existence of such an elusive boson would have to be independently confirmed through complementary probes. Although such a range for the boson mass is currently outside of the experimental reach for terrestrial laboratories, possible signatures that can be exploited in the future include astrometry~\cite{GRAVITY:2019tuf,Tsai:2021irw}, as well as GW emission from level transitions or boson annihilation to gravitons~\cite{Arvanitaki:2014wva,Baryakhtar:2017ngi,Brito:2020lup}. The launch of eLISA~\cite{Amaro-Seoane:2012vvq} will make the latter a particularly promising target for the boson mass window considered here. If independent probes confirm that Sgr A$^\star$'s spin is low, as suggested by Ref.~\cite{Fragione:2020khu}, this could hint to the existence of a light boson with $\mu \sim 10^{-17}\,{\rm eV}$ (conversely to how the putative high spin of M87$^\star$ is used to exclude a range of ultralight boson masses in Ref.~\cite{Davoudiasl:2019nlo}).

While the possibility of observing a superradiant phase at present time is small, it is not vanquished. Possibilities include a recent bosenova event that depends on nonlinear self-interacting terms, leading to the destruction of the cloud~\cite{2001PhRvL..86.4211R, 2001Natur.412..295D,Yoshino:2012kn}. Quantitative simulations of the bosenova process, affecting how much rotational energy falls back to the BH, vary from different types of self-interactions. In fact, for the case of a scalar field in the presence of quartic self-interactions, new processes emerge, such as the emission of both relativistic and nonrelativistic waves to infinity, and the excitation of forced oscillations which are eventually reabsorbed by the BH~\cite{Baryakhtar:2020gao}. A model featuring a Higgsed vector field coupled to a complex scalar field has also been recently explored~\cite{East:2022ppo}. In both cases, the occupation number of the cloud is limited by the presence of the new interaction and superradiance halts after extracting a maximum angular momentum $J_{\max}$, a fraction of which is lost through the emission of GWs and boson fields at infinity. Although this process has been simulated through lattice computations~\cite{Yoshino:2012kn, Yoshino:2013ofa, Yoshino:2015nsa}, more efforts are underway to better understand the conditions under which the bosenova events actually proceed in bursts, or whether the emission proceeds through a saturation of the cloud with a steady-state outflow. We leave a detailed study of these and related aspects to future work.

Another possibility is a time-dependent boson mass, related to a dark Higgs sector whose vacuum expectation value evolves over timescale longer than the superradiant one, triggering superradiance at specific times. Such a scenario would require that the boson mass $\mu = \mu(t)$ satisfies the following conditions: \textit{i)} superradiance occurs at present time $t_0$, $GM_0\mu(t_0) \sim \mathcal{O}(0.1)$, and \textit{ii)} the rate of the change in the boson mass be smaller than the superradiance timescale, $\dot\mu/\mu \lesssim \Gamma_{j\ell}$. One possibility in this sense entails considering a time-varying vacuum expectation value~\cite{Graham:2015cka}. Additional future possibilities include incorporating the role of self-interactions~\cite{Fukuda:2019ewf,Baryakhtar:2020gao,East:2022ppo} and couplings to SM or even dark sector particles~\cite{Ikeda:2018nhb,Boskovic:2018lkj,Caputo:2021efm,Cannizzaro:2022xyw}, the impact of multiple boson species or higher excitation modes, and the interplay between ultralight bosons and modified gravity effects~\cite{Odintsov:2019mlf,Odintsov:2019evb,Nojiri:2019riz,Odintsov:2020nwm,Nojiri:2020pqr,Odintsov:2020iui,Oikonomou:2020qah}. We plan to return to these points in future work.

\begin{acknowledgments}
We thank Richard Brito and Vitor Cardoso for useful comments and discussions that lead to this work. Y.C.\ is supported by the China Postdoctoral Science Foundation under Grants No.\ 2020T130661, No.\ 2020M680688, the International Postdoctoral Exchange Fellowship Program, and by the National Natural Science Foundation of China (NSFC) under Grants No.\ 12047557. R.R.\ is supported by the Shanghai Government Scholarship (SGS). S.V.\ is supported by the Isaac Newton Trust and the Kavli Foundation through a Newton-Kavli Fellowship, by a grant from the Foundation Blanceflor Boncompagni Ludovisi, n\'{e}e Bildt, and by a College Research Associateship at Homerton College, University of Cambridge.
\end{acknowledgments}

\newpage
\bibliography{ref.bib}

\end{document}